\documentclass[aps,prl,twocolumn,superscriptaddress,twocolumn]{revtex4}

\usepackage{graphicx,amsmath,amssymb,bm,subfigure}
\usepackage[toc,page]{appendix}
\usepackage{float}
\usepackage{caption}
\begin{document}

\title{Quantum Simulation of Molecular Collisions with Superconducting Qubits}

\author{Emily J. Pritchett}
\affiliation{Department of Physics and Astronomy, University of Georgia, Athens, GA 30602}
\author{ Colin Benjamin}
\affiliation{Department of Physics and Astronomy, University of Georgia, Athens, GA 30602}
\affiliation{Center for Simulational Physics, University of Georgia, Athens, GA 30602}
\author{Andrei Galiautdinov}
\affiliation{Department of Physics and Astronomy, University of Georgia, Athens, GA 30602}

\author{Michael R. Geller}
\affiliation{Department of Physics and Astronomy, University of Georgia, Athens, GA 30602}
\author{Andrew T. Sornborger}
\affiliation{Department of Mathematics and Faculty of Engineering, University of Georgia, Athens, GA 30602}
\author{Phillip C. Stancil}
\affiliation{Department of Physics and Astronomy, University of Georgia, Athens, GA 30602}
\affiliation{Center for Simulational Physics, University of Georgia, Athens, GA 30602}
\author{John M. Martinis}
\affiliation{Department of Physics, University of California at Santa Barbara, Santa Barbara, California 93106}
\date{July 8, 2010}

\begin{abstract}
We introduce a protocol for the fast simulation of $n$-dimensional quantum systems on $n$-qubit quantum computers with tunable couplings.  A mapping is given between the control parameters of the quantum computer and the matrix elements of  $H_{\rm s}(t)$, an arbitrary, real, time-dependent $n\times n$ dimensional Hamiltonian that is simulated in the $n$-dimensional `single excitation' subspace of the quantum computer.   A time-dependent energy/time rescaling minimizes the simulation time on hardware having a fixed coherence time.  We demonstrate how three tunably coupled phase qubits simulate a three-channel molecular collision using this protocol, then study the simulation's fidelity as a function of total simulation time.

\end{abstract}
\maketitle

A quantum computer can significantly reduce the resources necessary to simulate quantum mechanical systems \cite{Feynman}.  Typically, quantum simulation algorithms algorithms  construct the simulated system's time evolution operator, energies and/or  eigenstates from a universal set of gates \cite{Lloyd,DQS,operator ordering,Kassal,photonic DQS,NMR DQS,Limitations}.  Alternatively, ultracold atoms, trapped ions, and liquid-state NMR have directly emulated the time evolution of certain other quantum systems \cite{cold atoms, trapped ions, NMR}.  Recent experimental progress suggests that quantum simulation will be one of the first practical applications of quantum computation  \cite{photonic DQS, NMR DQS, cold atoms, trapped ions, NMR, Nori}.  

In principle, an $n$-qubit quantum computer can store the state of  any $N=2^n$ dimensional quantum system,  an exponential reduction in the resources necessary to store quantum information on a classical computer.  However, simulation may require $\sim N^2=2^{2n}$ elementary gates per time step unless the simulated Hamiltonian has special properties,  e.g. locality \cite{Barenco,Lloyd}.  Even for these special  Hamiltonians, fully digital quantum simulation often requires an excessive number of gates for  current quantum computing technology \cite{Limitations,Kassal}.  

 In this Letter, we show that a subspace of a tunable $n$-qubit quantum computer can emulate an arbitrary $n$-dimensional quantum system, trading an exponential reduction in resources for simulations of a wider variety of Hamiltonians.  This subspace simulates other quantum systems very different from the computer itself in an amount of time that is independent of $n$.  By comparison, classical simulation of an $n$-dimensional quantum system requires $\sim n^3$ elementary operations per time step.  While the most efficient quantum simulation algorithms offer an exponential reduction in both qubits and elementary operations, they typically apply to specific, fundamental time-independent Hamiltonians, or those already similar to that of the computer itself.  We show that with a more modest polynomial reduction in resources, a subspace of a tunable quantum computer can simulate any real, time-dependent Hamiltonian.
 
We begin by outlining the theory behind our approach to simulation.  First, we identify an $n$-dimensional invariant subspace suitable for quantum simulation.  Then we define a time dependent energy/time rescaling that maximizes the speed of the simulation within the constraints of the quantum computer.  Finally, the control parameters of the quantum computer are given explicitly  as a function of the matrix elements of $H_{\rm s}(t)$.

Our approach is tested by performing a simulation of a molecular collision with a circuit of tunably coupled Josephson phase qubits. 
Molecular collisions and electronic structure calculations are widely studied as important applications of quantum simulation techniques \cite{Kassal, NMR DQS, photonic DQS}. We show in detail how a superconducting circuit of three tunably coupled Josephson phase qubits simulates a three channel Na-He collision.   Finally, we discuss the relationship between simulation fidelity and total simulation time for this particular example. 

{\it An n-Dimensional Subspace} of the full quantum computer's Hilbert space, ${\cal H}$, can emulate another quantum system at all times only if it is invariant to the time evolution generated by the computer's Hamiltonian $H_{\rm qc}$ (so that the subspace is well-isolated from the rest of ${\cal H}$ and evolves unitarily).  We model $H_{\rm qc}$ as 
\begin{eqnarray}
H_{\rm qc}(t)=\sum_{i=1}^n-{\epsilon_i(t)\over{2}}\sigma_i^z+{1\over{2}}\sum_{i\neq j}g_{ij}(t)J_{\mu\nu}\sigma^\mu_i\otimes\sigma^\nu_j,
\end{eqnarray}
 where $\epsilon_i(t)$ are the uncoupled qubit energies, $g_{ij}(t)=g_{ji}(t)$ are the pairwise qubit interaction strengths, $J_{\mu\nu}$ gives the relative size of the $\sigma_i^\mu\otimes\sigma_j^\nu$ interaction, and $\mu,\nu\in\{0,x,y,z\}$ are summed over.  While $\epsilon_i(t)$ and $g_{ij}(t)$ may in general be time-dependent, the time-independent structure of qubit interaction is specified by $J_{\mu\nu}$, a dimensionless tensor that is typically fixed by a given architecture and is identical between each pair of qubits. In the weak coupling limit, $|g_{ij}| ||J_{\mu\nu}||/\epsilon_i\ll 1$, subspaces of ${\cal H}$ are invariant to time evolution generated by $H_{\rm qc}$ if spanned by computational basis states having the same number of excited (tunable) qubits.
The 	`single excitation subspace', denoted as ${\cal H}_n$, is an $n$-dimensional invariant subspace spanned by $ |i\big>_{n} \equiv|00..01_i..0_n\big>$ for all $i=1, 2 ...,n$.
 
 The control parameters $\epsilon_i(t)$ and $g_{ij}(t)$ directly control the Hamiltonian that ${\cal H}_n$ simulates.  We define $H_n$ as 
$H_{\rm qc}$ projected into the single excitation subspace, 
 \begin{equation}
 H_{n}(t)\equiv P H_{\rm qc}(t)P^\dagger
 \end{equation}
 where $P$ is an  $n\times 2^n$ dimensional operator that projects ${\cal H}$ onto ${\cal H}_n$.  Up to an additive energy shift, $H_n$ has matrix elements 
 \begin{eqnarray}
H_{ n}^{ij}(t)&\equiv&
\begin{cases}
\epsilon_i(t)-{\alpha}\sum_{k\neq i}g_{ik}(t), \  i=j\\
g_{ij}(t) ,  \ i\neq j
\end{cases}
\end{eqnarray}
with  $\alpha\equiv2(J_{zo}+J_{zz})$.  We assume $J_{xx}+J_{yy}\neq 0$ and normalize $J_{\mu\nu}$ so that $J_{xx}+J_{yy}=1$.  In the weak coupling limit, ${\cal H}_n$ is approximately invariant and generated by $H_n$:
\begin{eqnarray}
U_n(t)&\equiv& PU_{\rm qc}(t)P^\dagger\nonumber\\
&\simeq&{\cal T}e^{-{i\over{\hbar}}\int_{0}^{t_{\rm }}H_{n}(t_{\rm }')dt_{\rm }'}
\end{eqnarray} 
where ${\cal T}$ is the time-ordering operator.
$H_n$ generates $U_n$ exactly when no matrix elements of $H_{\rm qc}$  mix ${\cal H}_n$ with the rest of ${\cal H}$ (i.e. $J_{0x}=J_{0y}=J_{zx}=J_{zy}=0$). 
 The $(n^2 +n)/2$ parameters $\epsilon_i(t)$ and $g_{ij}(t)$ independently control each of the $(n^2+n)/2$ matrix elements of the real $H_n$ and can therefore be used to simulate any arbitrary, real Hamiltonian in ${\cal H}_n$.

While we can simulate $H_s$ in ${\cal H}_n$ by choosing $\epsilon_i(t)$ and $g_{ij}(t)$ so that $H_{n}(t)=H_{\rm s}(t)$ for all $t$,  a direct mapping between Hamiltonians  limits the computer to simulating other quantum systems with similar energy scales over lengths of time within the computer's coherence time.  
Fortunately, simulation of $H_{\rm s}$ only requires equality up to an overall phase between $U_n$ and the time evolution operator generated by $H_{\rm s}$:
\begin{eqnarray}
U(t)&\equiv&{\cal T}e^{-{i\over{\hbar}}\int_{t_{\rm i}}^{t}H_{\rm s}(t')dt'}\nonumber\\
&=&e^{i\phi(t)} U_n(t_{\rm qc}(t)).
\label{eqn:TEOequality}
\end{eqnarray}
The time elapsed on the quantum computer, $t_{\rm qc}(t)$, is a strictly increasing function of simulated time $t$,
admitting a much less restrictive relationship between Hamiltonians:
\begin{eqnarray}
H_{\rm s}(t) + c(t)=\lambda(t)H_{n}(t_{\rm qc}(t)).
\end{eqnarray}
 $c(t)$ is a time-dependent, additive energy shift giving the overall phase difference $\phi(t)={1\over{\hbar}}\int_{t_{\rm i}}^tc(t')dt'$, and we have introduced a positive, time-dependent energy/time scaling
 \begin{eqnarray}
\lambda(t)\equiv dt_{\rm qc}/dt.
\end{eqnarray}

{\it The energy/time scaling} $\lambda(t)$ determines the speed of the simulation.  By carefully minimizing  $\lambda(t)$, we reduce the total simulation time and, consequently, the error due to decoherence.  $\lambda(t)$ is bounded from below by experimental constraints on the allowed values of control parameters $\epsilon_i(t)$ and $g_{ij}(t)$ as well as their maximum rates of change.  Suppose qubit interaction strengths can vary in a range $g_{ij}(t)\in[-g_{\rm max},g_{\rm max}]$, and the uncoupled qubit energies can vary in a range  $\epsilon_i(t)\in[\epsilon_{\rm min},\epsilon_{\rm max}]$.  
For convenience, we define a simulated energy $E_i(t)$ analogous to $\epsilon_i(t)$ when diagonal contributions from qubit interactions are anticipated:
  \begin{eqnarray}
E_i(t)&\equiv&H_{\rm s}^{ii}(t)+\alpha\sum_{j\neq i}H_{\rm s}^{ij}(t).
\end{eqnarray}
Using this definition together with equations (3) and (6), we relate the control parameters of the quantum computer to the simulated energies in $H_{\rm s}(t)$:
\begin{eqnarray}
g_{ij}(t)&=&H_{\rm s}^{ij}(t)/\lambda(t)\nonumber\\
\epsilon_i(t)&=&[E_i(t)-c(t)]/\lambda(t).
\end{eqnarray}
 By choosing $c(t)=E_{\rm max}(t)-\lambda(t)\epsilon_{\rm max}$ where $E_{\rm max}(t)$ is the largest  value obtained by the $E_j(t)$ at a particular $t$, we force each $\epsilon_i$ to be as large as possible and therefore minimize leakage out of ${\cal H}_n$.

 Each of the computer's control parameters remains within its allowed range when $\lambda(t)$ is larger  than $(n^2 + n)/2$ energy ratios at all times:
  \begin{eqnarray}
\lambda(t)& \geq&
\begin{cases}
|H_{\rm s}^{ij}(t)|/{g_{\rm max}}, \  i \neq j\\
{\Delta E_i(t)/{\Delta \epsilon_{\rm max}}}
 \end{cases}
\end{eqnarray}
where $\Delta E_i(t)\equiv E_{\rm max}(t)-E_i(t)$ and 
 $\Delta\epsilon\equiv \epsilon_{\rm max}-\epsilon_{\rm min}$.
 $\lambda(t)$ is also bounded by
constraints on the speeds with which control parameters can change.
 Suppose $v^\epsilon_i(t_{\rm qc})\equiv d\epsilon_i(t_{\rm qc})/dt_{\rm qc}$ and $v_{ij}^g(t_{\rm qc})\equiv d g_{ij}(t_{\rm qc})/dt_{\rm qc}$ can never be larger in magnitude than $v^\epsilon_{\rm max}$ and $ v^g_{\rm max}$ respectively.  Then for all  $t$,
\begin{eqnarray}
v^g_{\rm max}&\geq&{1\over{\lambda^2}}\left|{dH_{\rm s}^{ij}(t)\over{dt}}-{H_{\rm s}^{ij}(t)\over{\lambda}}{d\lambda\over{dt}}\right|
\end{eqnarray}
(and similarly for $v^\epsilon_{\rm max}$).

{\it To simulate $H_s(t)$ in ${\cal H}_n$},  we first choose $\lambda(t)$ as small as both inequalities (10) and (11) allow, guaranteeing a fast simulation within the experimental constraints of the quantum computer.  We integrate over $\lambda(t)$ to calculate $t_{\rm qc}$ as a function of $t$:
\begin{equation}
t_{\rm qc}(t)=\int_{t_{\rm i}}^{t}\lambda(t') dt' + t_{\rm qc}(t_{\rm i}).
\end{equation}
With both $\lambda(t)$ and $t_{\rm qc}(t)$ known, we can explicitly map the matrix elements of $H_{\rm s}$ to the control parameters of the quantum computer:
\begin{eqnarray}
\epsilon_i(t_{\rm qc}(t))&=&\epsilon_{\rm max} + \Delta E_i(t)/\lambda(t) \nonumber\\
g_{ij}(t_{\rm qc}(t))&=& H_{\rm s}^{ij}(t)/\lambda(t).
\end{eqnarray}

 {\it To demonstrate our theory in detail}, we describe three Josephson phase qubits simulating a three-channel collision between a sodium and a helium atom.  For three phase qubits with tunable inductive coupling,
\begin{eqnarray}
H_{\rm qc}(t)&=&\sum_{i=1}^{3}-{\epsilon_i(t)\over{2}}\sigma_i^z+{1\over{2}}\sum_{i\neq j}g_{ij}(t)\hat{\Phi}_i\otimes \hat{\Phi}_j
 \end{eqnarray}
where $\hat{\Phi}_i$ is defined in terms of the matrix elements  $\varphi_{jk}=\big<j|\hat{\varphi}_i|k\big>$ of the local Josephson phase operator in the  computational basis of the $i$th qubit:
\begin{eqnarray}
\hat{\Phi}_i
&\equiv&\sigma_i^x+{ {\varphi_{00}-\varphi_{11}\over{2\varphi_{01}}}}\sigma^z_i +{{\varphi_{11}+\varphi_{00}\over{2\varphi_{01}}}}\sigma^0_i.
\end{eqnarray}
Both the $\epsilon_i$ and the $\varphi_{jk}$ depend on $\Phi_x$, the externally applied flux through the superconducting circuit.  External flux bias is quantified by a dimensionless parameter $s_i(t)=\Phi_x/\Phi_x^*$ where  $\Phi_x^*$ is the qubit's critical flux bias, or alternatively, by the dimensionless well depth $\Delta U/\hbar\omega_p$   \cite{Martinis}.   We consider external bias  values for which $s\in[.89,.90]$ and $\Delta U/\hbar\omega_p\in[13.7,15.5]$.  In this range, $\Delta \epsilon/ h=190 {\rm MHz}$ while $\hat{\Phi}_i\simeq \sigma_i^1 + 11\sigma_i^0$ varies little.  A tunable mutual inductance independently controls the couplings $g_{ij}(t)$ between each pair of qubits.  We have assumed Josephson junction parameters $I_0=2.93 \ \mu {\rm A}$, $C=1.52\  {\rm pF}$,  and $L=808 \ {\rm pH}$. 

 An $n$-dimensional subspace can simulate a molecular collision only after we project the full, many-body Hamiltonian of the interacting electrons and nuclei into an $n$-dimensional basis.  We construct the collision Hamiltonian from Born-Oppenheimer energies and nonadiabatic  couplings calculated previously for three molecular channels:  Na($3s$) + He($1s^2$) [$1~^2\Sigma^+$] and Na($3p$) + He($1s^2$)[$1~^2\Pi$; $2~^2\Sigma^+$]  \cite{lin}, labeled as $|1\big>_{\rm s}$, $|2\big>_{\rm s}$ and $|3\big>_{\rm s}$ respectively.  The energies are stored for fixed values of the internuclear distance $R$, which we assume takes straight-line trajectories in a standard semiclassical approximation:  $R(t)=\sqrt{b^2 + v^2t^2}$ where $v$ is the incoming particle's velocity and $b$ is the impact parameter of the collision.  

Figure 1 outlines our simulation protocol for $H_{\rm s}(t)$ describing a three-channel Na-He collision.  The matrix elements of $H_{\rm s}(t)$ are displayed in Fig. 1(a) for a given semiclassical trajectory $R(t)$.  Directly below, we plot the energy/time scaling parameter $\lambda(t)$ as a black curve enveloping the six energy ratios given in Eq. (10).  A small $\lambda(t)$ speeds the quantum computer through times when the internuclear distance $R$ is large, but as $R$ decreases ($t\rightarrow 0$), a relatively small $g_{\rm max}$ value constrains the growing couplings.   $\lambda(t)$ increases over two orders of magnitude, creating a highly nonlinear relationship between $t_{\rm qc}$ and $t$, as shown in Fig. 1(c).  This effectively stretches the  portion of the collision when internuclear distance is small over the entire simulation, as can be seen in the plot of the quantum computer's control parameters as a function of $t_{\rm qc}$ in Fig. 1(d).

\begin{figure}[!]

 {\includegraphics[width=0.48\textwidth]{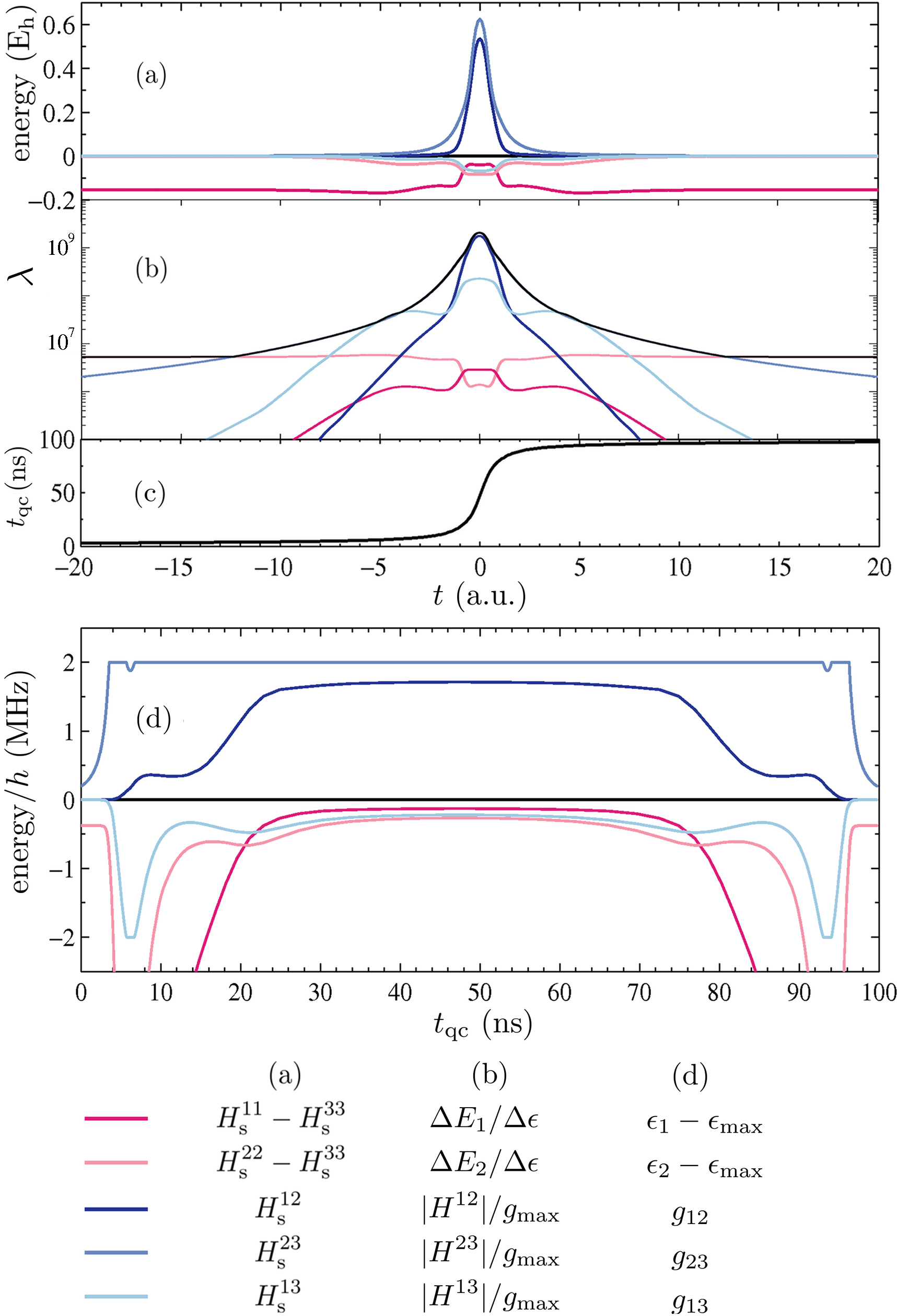}}

\vspace{-.1 cm}

  \caption{ 
  (color online)  $H_{\rm s}(t)$ describes a three channel Na-He collision with $b=0.5$ and $v=1.0$.  (a) Matrix elements of $H_{\rm s}$ as a function of time in atomic units (${\rm E_ h} = 27.21$ eV and the atomic unit of time is $2.419\times 10^{-8}$ ns).
 (b) The dimensionless time scaling parameter $\lambda(t)$ envelopes six energy ratios ($\Delta E_3=0$ for all $t$).  We assume $g_{\rm max}/h=2.0 {\ \rm   MHz}$ and  $\Delta\epsilon_{\rm max}/h={\rm 190 \  MHz}$. (c)  Plot of $t_{\rm qc}(t)$ for the case of $t_{\rm qc}(t_{\rm i})=0$, $t_{\rm i}=-40\ {\rm a.u.}$.  (d) Control parameters that simulate $H_{\rm s}(t)$ plotted as a function of $t_{\rm qc}$ ($\epsilon_3=\epsilon_{\rm max}$ for all $t_{\rm qc}$).   }
\vspace{-.5 cm}

\end{figure}

{\it To study the fidelity of the simulation}, we compare the exact and simulated time evolution operators, $U(t)$ and $U_n(t_{\rm qc}(t))$ respectively, by plotting (in Fig. (2)) transition probabilities out of $|1\big>$:
\begin{equation}
P_{1i}(t)\equiv|\big<i|U(t)|1\big>|^2.
\end{equation}
\noindent Because the exact transition probabilities evolve differently with $t$ than the simulated evolve with $t_{\rm qc}$, we define a time-dependent transition fidelity which accounts for time scaling,
\begin{equation}
F(t)\equiv|{\vphantom{\big>}}_{\rm s}\big<1|U^\dagger (t) U_n(t_{\rm qc}(t))|1\big>_{n}|^2,
\end{equation}
and a time-dependent leakage out of ${\cal H}_n$, 
\begin{equation}
 L(t)\equiv\textstyle \sum_{{_\perp}}|\vphantom{\big>}_{\perp}\! \big<i|U_{\rm qc}(t_{\rm qc}(t))|1\big>_{n}|^2
 \end{equation}
where $\sum_\perp$ is the sum over all computational basis states $|i\big>_{\perp}$ orthogonal to ${\cal H}_{\rm n}$.  In the upper part of  Fig. (3), fidelity and leakage are plotted together for four different $g_{\rm max}$ values. 
\begin{figure}[t]          
{\includegraphics[width=0.48\textwidth]{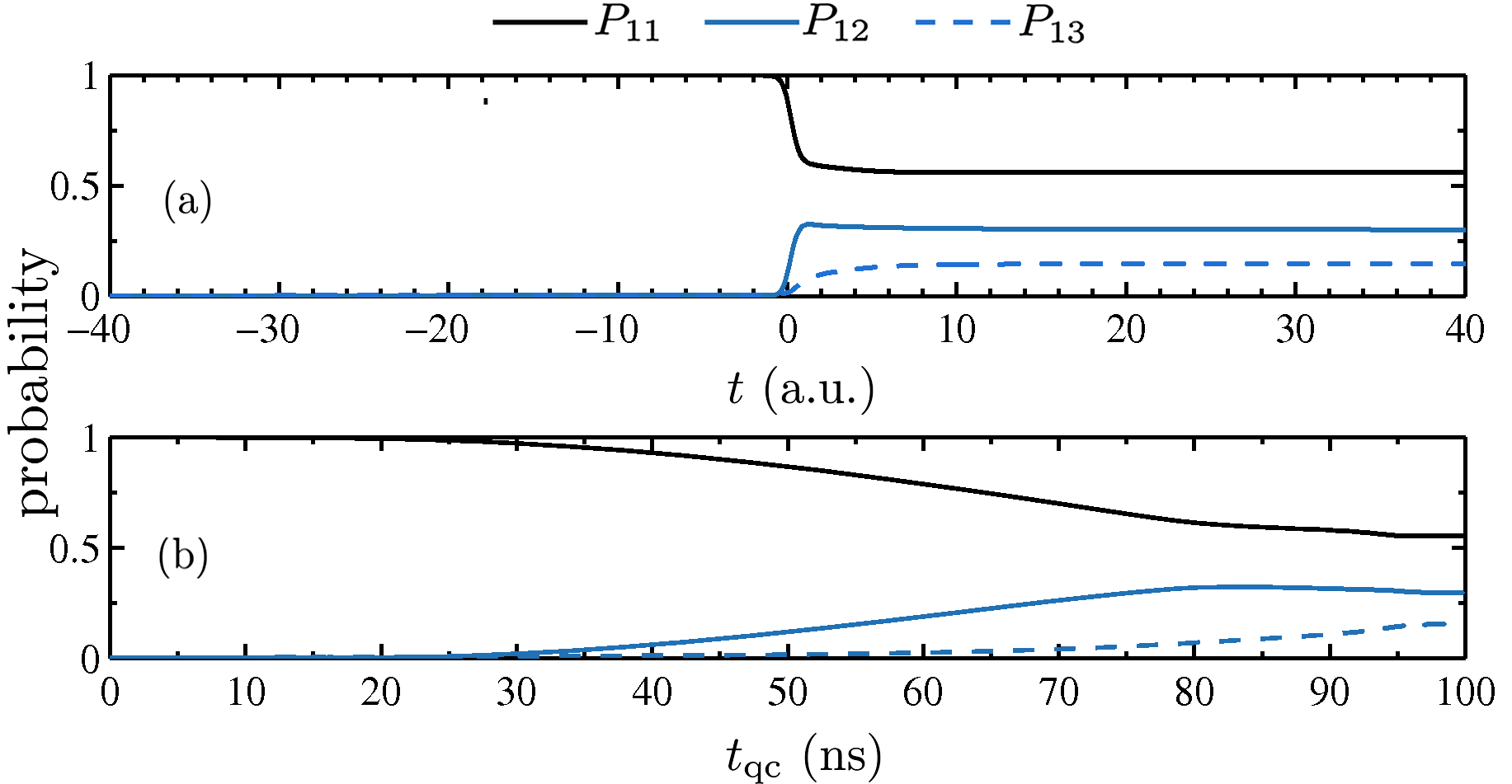}}
   \vspace{-.65 cm}
  \caption{(a) (color online) Exact transition probabilities generated by  $H_{\rm s}(t)$ shown in Fig. 1(a).  (b) Transition probabilities simulated with parameter profiles given in Fig. 1(d).  Final simulation fidelity is 0.998.}   
  \vspace{-.55 cm}
\end{figure}

Minimizing $g_{\rm max}||J_{\mu\nu}||/\epsilon_{\rm min}$, either by decreasing $g_{\rm max}$ or by increasing $\epsilon_{\rm min}$, reduces leakage and thus improves simulation fidelity.
   In this example, we find simulation fidelity more sensitive to the cutoff in $g_{\rm max}$ because leakage is most prominent when the interatomic distances are small ($t\rightarrow 0$) and the diabatic couplings between channels are the dominant terms.  By reducing $g_{\rm max}$ we also increase $\lambda(t)$ and thus the total simulation time, as studied in the lower plot of Fig. (3).    To increase fidelity from .9990  to .9999 we need to increase the simulation time by a factor of $~\sim3$, a relationship that is independent of $n$.  While not introducing specific models of decoherence, we note that high fidelity simulations are possible on superconducting qubits with coherence times around $100$ ns.



%
When applied to molecular collisions, our approach to quantum simulation requires classical overhead to 
project the fundamental, time-independent, many-body Hamiltonian into an $R$-dependent, $n$-channel $H_{\rm s}$.  The quantities of physical interest, cross sections, are obtained by integrating the final transition probabilities over many semiclassical trajectories with different impact parameters, which requires no further classical overhead.    A classical simulation of transition probabilities requires $\sim n^3$ elementary operations per time step for a single impact parameter, thus cross section calculations are computationally intensive for large $n$.  
Alternatively,  simulation time is independent of $n$ using our protocol, so once the $R$-dependent $H_{\rm s}$ has been calculated, cross sections can be obtained quickly.

\begin{figure}[t]
{\includegraphics[width=0.48\textwidth]{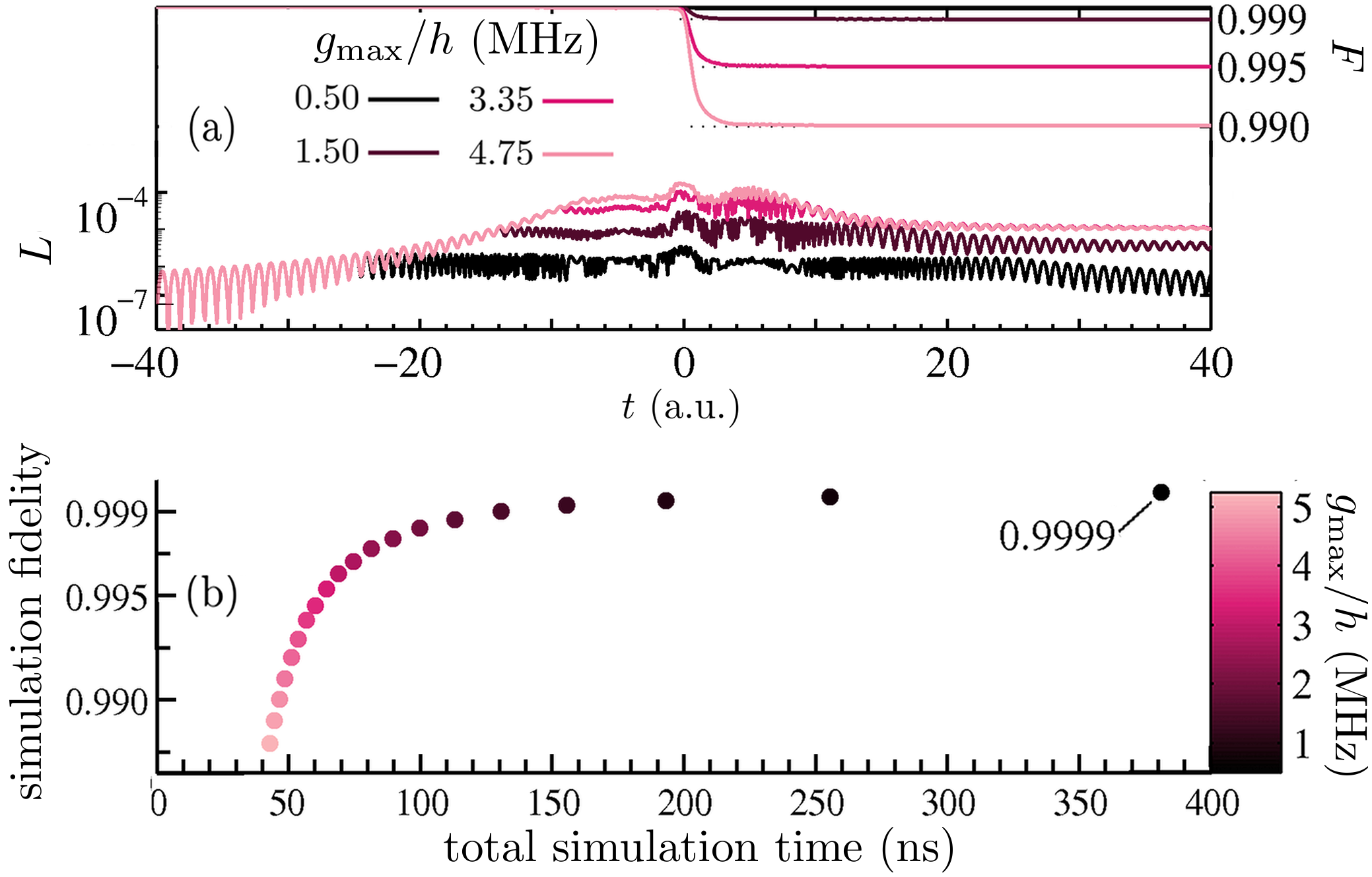}}
\vspace{-.45 cm}
   \caption{ (color online) (a) Fidelity and leakage as a function of simulation time for four different $g_{\rm max}$ values, with all other parameters the same as in Fig. 1. (b) Final simulation fidelity versus total simulation time for  varying $g_{\rm max}$.  The $g_{\rm max}$ value is referenced by the shade of the data point.}
  \label{fig:Mapping} 
  \vspace{-.45 cm}
\end{figure}

In summary, we have presented a straightforward protocol for quantum simulation that can be implemented with currently available superconducting quantum computing technology.  While a promising application of quantum computation, current  quantum simulation protocols require a threshold number of gates and qubits that prohibits fully digital quantum simulations from being demonstrated on available quantum computers.  However, we have shown how quantum computers of only a few qubits can simulate arbitrary quantum systems accurately and quickly even before they reach the regime of fault tolerant quantum computation.

\begin{acknowledgements}
It is a pleasure to thank Joydip Ghosh for interesting discussions.  
This work was partially supported by NSF grants PHYS-0939849 and
PHYS-0939853 from the Physics at the Information Frontier Program.

\end{acknowledgements}

\end{document}